\documentclass[referee]{raa}            

\usepackage{graphicx,times}             
\usepackage{textcomp}
\usepackage{amsmath}
\usepackage{amssymb}{\tiny }
\begin{document}
\title{Cosmological Parameters For Spatially Flat Dust Filled Universe In Brans-Dicke Theory }



   \volnopage{Vol.0 (200x) No.0, 000--000}      
   \setcounter{page}{1}          
\author{G. K. Goswami}
\institute{ Department of Mathematics, Kalyan P. G. College, Bhilai-490 006, C. G., India;\\ {\it gk.goswam9@gmail.com}}
\date{Received~~2009 month day; accepted~~2009~~month day}
\abstract{ In this paper, we have investigated  late time acceleration for a spatially flat  dust filled  Universe in Brans-Dicke theory in the presence of a positive cosmological constant $\Lambda$ . Expressions for Hubble's constant, luminosity distance and apparent magnitude have been obtained for our model. The theoretical results are compared with the observed values  of the  the latest 287 high red shift ($ .3 \leq z \leq 1.4$ ) SN Ia supernova data's  taken from Union 2.1 compilation to estimate the present values of the matter and dark energy parameters $(\Omega_{m})_0$ and $(\Omega_{\Lambda})_0$. We have also estimated the present value of Hubble's constant $H_0$ in the light of a updated sample of Hubble parameter measurements including 19 independent data points.  The results are found to be  in good agreement with recent astrophysical observations. We have also calculated various physical parameters such as the matter and dark energy densities, the present age of the universe and deceleration parameter. The value for BD-coupling constant $\omega$ is set to be 40000 on the basis of accuracy of the solar system tests and  recent experimental evidence.
}

   \authorrunning{G K Goswami }            
   \titlerunning{Cosmological Parameters For Dust Filled Universe In Brans-Dicke Theory }  
 \maketitle

%
%
\section{Introduction}           
\label{sect:intro}
The two independent groups headed by  Riess and  Perlmutter via type Ia supernovae (SNeIa ) (Riess at el ~\cite{ref1}; Perlmutter at el~\cite{ ref2}), found that our universe is accelerating at present. Several theories have been put forward  to explain this remarkable discovery
(Spergel et al.~\cite{ref3}; Bennett et al.~\cite{ref4};Tegmark et al.~\cite{ref5}).
An exotic bizarre form of the energy called as dark energy is proposed to understand the accelerating expansion.
The dark energy is expected to possess a negative pressure, which detracts matters from each other and creates
acceleration in the universe. The simplest candidate of dark energy is the positive cosmological constant $\Lambda$  which is
considered as a source  with equation of state ~$p_\Lambda=-\rho_\Lambda$. The standard Friedmann
Robertson Walker ( FRW) model of the universe with cosmological constant as a source of dark energy is often known as
$\Lambda$-CDM cosmological model ( Copeland at el.~ \cite{ref6};  Gron and  Hervik~\cite{ ref7}).
Basically, the standard FRW model represents decelerating universe but presence of cosmological constant
as a source and its specific value makes the model accelerating. It is found that the $\Lambda$-CDM
model is in good agreement with  latest observations (Abazajian et al. \cite{ref8}; Sahni and  Starobinsky \cite{ref9}).
Recently  Goswami at el. ( \cite {ref10};\cite {ref11};\cite {ref12}) have developed  $\Lambda$-CDM type
models  for Bianch type I anisotropic universe. \\
Apart from $\Lambda$-CDM cosmological model, alternative explanations for the
accelerated expansion in terms of scalar fields like
quintessence ( Caldwell at el~\cite{ref13}) , K-essence ( Chiba at el.~ \cite{ref14}) , phantom fields (Caldwell~ \cite{ref15})
and Chaplin gas (Kamenshchik at el.~ \cite{ref16})  are available, a number of
models involving the cosmological term, especially
time-varying Λ have been proposed ( Carvalho at el.~\cite{ref17}; Wetterich~\cite{ref18}; Arbab~\cite{ref19}; Padmanabhan ~\cite{ref20}; Vishwakarma~\cite{ref21}; Shapiro and  Sola~\cite{ref22};  Dutta Choudhury and  Sil~ \cite{ref23}).\\
 It is worth to investigate  effect of cosmological ‘constant’ in
the Brans-Dicke theory of gravity ( Brans and Dicke~ \cite{ref24}) which  describe  evolution of the universe that  explain
accelerating phase of expansion  in the current epoch. In a recent paper,  Hrycyna and  lowski~ (\cite{ref25}) compared dynamical evolution of the standard cosmological model from the view of both  Brans-Dicke and general theory of relativity. Singh and Singh~ (\cite{ref26}) investigated a cosmological model in Brans-Dicke theory by considering cosmological “constant” as function of scalar field $\phi$.  Pimentel~ (\cite{ref27})
obtained exact cosmological solutions in Brans-Dicke theory with uniform
cosmological “constant”. A class of flat FRW cosmological models with
cosmological “constant” in Brans-Dicke theory have also been obtained by Azar
and Riazi~( \cite{ref28}). The age of the universe from a view point of the nucleosynthesis
with Λ term in Brans-Dicke theory was investigated by Etoh et al~ (\cite{ref29}). Azad and
Islam~( \cite{ref30}) extended the idea of Singh and Singh~( \cite{ref26}) to study cosmological constant in Bianchi type I modified Brans-Dicke cosmology. Recently Qiang et al~( \cite{ref31}) discussed cosmic acceleration in five dimensional Brans-Dicke theory using interacting Higgs and Brans-Dicke fields. Smolyakov~( \cite{ref32}) investigated a model which provides the necessary value of effective cosmological “constant” at the
classical level. Embedding general relativity with varying cosmological term in
five dimensional Brans-Dicke theory of gravity in vacuum has been discussed by
Reyes et al~( \cite{ref33}).\\
In this paper, we have investigated  late time acceleration for a spatially flat  dust filled  Universe in Brans-Dicke theory in the presence of a positive cosmological constant $\Lambda$ . The paper is organized as follows: In section 2, The BD-field equations are developed for a  universe filled
with cosmic fluid  as source of matter in spatially homogeneous and isotropic space-time . In Section 3, we have obtained expression for Gravitational constant in term of red shift by solving BD-field equations. The value for Coupling constant $\omega$ is set to be 40000 on the basis of accuracy of the solar system tests and recent experimental evidence. In this section, we have also obtained expression for Hubble's constant and relation ship between energy parameters $\Omega_m $ and $\Omega_{\Lambda}$. In section 4, expressions for Luminosity distance and apparent magnitude have been obtained . The estimation of   energy parameters and  Hubble's constant at present are dealt in the Section 5 and 6. In Section 7, we have obtained various
 physical parameters such as the matter and dark energy densities,  present age of the universe and value of deceleration parameter  on the basis of  values of $(\Omega_{m})_0$, $(\Omega_{\Lambda})_0$ \& $ H_0 $ obtained by us. The model predicts that   acceleration in the  universe  had begun in the
 past at $z_{c}=0.6818 \thicksim 7.2371\times10^9 yrs$ before from present. Finally the conclusions of the paper are presented in section 8. The results of our investigation are consistence with the latest large scale structure measurements by surveys like BOSS, wiggleZ and BAO,
 and WMAP or Planck results for CMB anisotropies ( Ade et al.~ \cite{ref34}; Ade et al.~ \cite{ref35}; Aubourg at el~\cite{ref36}; Anderson at el~
\cite{ref37}; Delubac et al.~\cite{ref38}; Blake et al. ~\cite{ref39}).
 WMAP quoted the value of dark energy densities, $\Omega_{\Lambda}$ = $0.7184$,
 where as the combined WMAP+CMB+BAO+BOSS surveys put $\Omega_{\Lambda}$ =  $0.7181$.
 We have obtained  $\Omega_{\Lambda}$ = $0.704$.
\section{BD-Field Equation}
BD field equations are obtained from following action
\begin{equation}\label{eq1}
 \delta \int \sqrt{-g} \left\lbrace \phi \left(  R-2\Lambda\right)+\omega\frac{\phi_{k}\phi^{k}}{\phi} \right\rbrace
\end{equation}
The field equations are
\begin{equation}\label{eq2}
R_{ij}-\frac{1}{2}R + \Lambda g_{ij}=- \frac{8\pi}{\phi c^4}T_{ij}-\frac{\omega}{\phi^2}\left( \phi_i \phi_j -\frac{1}{2}g_{ij}\phi_k \phi^k\right) -\frac{1}{\phi}\left( \phi_{i;j} - g_{ij} \Box \phi \right)
\end{equation}
\begin{equation}\label{eq3}
(2\omega+3)\Box \phi=  \frac{8\pi  T}{ c^4}+2\Lambda \phi
\end{equation}
We consider FRW spatially
homogeneous and isotropic  space-time given by
\begin{equation}\label{eq4}
ds{}^{2}=c^{2}dt{}^{2}-a(t){}^{2}[dr{}^{2}/(1+kr^{2})+r^{2}({d\theta{}^{2}+sin{}^{2}\theta
	d\phi{}^{2}})]
\end{equation}
Where k=-1 for closed universe, k=1 for open universe and k=0 for
spatially flat universe. $a(t)$ is scale factor. \\
The energy momentum tensor is taken as that of  a perfect fluid. This is given by
\begin{equation}\label{eq5}
T_{ij}=(p+\rho)u_{i}u_{j}-pg_{ij},
\end{equation}
Where $g_{ij}u^{i}u^{j}=1$ and $u^{i}$ is the 4-velocity vector.\\
In co-moving co-ordinates
\begin{equation}\label{eq6}
u^{\alpha}=0,~~~~~~~~~\alpha=1,2, 3.
\end{equation}
The field equations (\ref{eq2}), for the line element (\ref{eq4}), are obtained as
\begin{equation}\label{eq7}
2\frac{\ddot{a}}{a}+\frac{\dot{a}^2}{a^2} + \frac{\omega \dot\phi^2}{2\phi^2}+2\frac{\dot{\phi}}{\phi}\frac{\dot{a}}{a} +\frac{\ddot{\phi}}{\phi}= -\frac{8\pi }{\phi c^{2}} p+\frac{kc^{2}}{a^{2}}  +\Lambda c^2
\end{equation}
\begin{equation}\label{eq8}
\frac{\dot{a}^2}{a^2}+\frac{\dot{\phi}}{\phi}\frac{\dot{a}}{a}- \frac{\omega \dot\phi^2}{6\phi^2} = \frac{8\pi }{3\phi c^{2}}\rho+\frac{kc^{2}}{a^{2}} +\frac{\Lambda c^2}{3}.
\end{equation}
\begin{equation}\label{eq9}
\frac{\ddot{\phi}}{\phi}+3 \frac{\dot{\phi}}{\phi}\frac{\dot{a}}{a}= \frac{8 \pi (\rho-3p)}{(2\omega+3)c^2\phi}+ \frac{2\Lambda c^2}{2\omega+3}
\end{equation}
\begin{equation}\label{eq10}
  \frac{\dot{\rho}}{\rho}+3\gamma\frac{\dot{(a)}}{a}=0
\end{equation}
\begin{equation}\label{eq11}
 \Lambda c^2+ 3\frac{kc^2}{a^2}= 3\frac{\ddot{a}}{a}+ 3\frac{\dot{a}^2}{a^2}-\omega \frac{\ddot{\phi}}{\phi}- 3\omega \frac{\dot{\phi}}{\phi}\frac{\dot{a}}{a}+ \frac{\omega \dot\phi^2}{2\phi}
 \end{equation}
 Where $\gamma$ is equation of state. $\gamma=1$ for dust dominated universe and  $\gamma=4/3$ for radiation filled universe.
\subsection{Dust Model}
The universe is (as at present)  dust-dominated, so we take   $p=0$. We define density parameters as
\begin{equation}\label{eq12}
\Omega_m=\frac{8\pi\rho}{3c^2H^2 \phi},\;\; \Omega_\Lambda=\frac{\Lambda c^2}{3H^2},
 and \;\;  \Omega_k=\frac{k c^2}{H^2a^2}
\end{equation}
We also define  decelerating parameter for scale factor $a$ as
  $$q= -\frac{\ddot{a}}{aH^2}$$
Equations (\ref{eq7}) to (\ref{eq9})  and (\ref{eq11}), then  become
 \begin{equation}\label{eq13}
 \Omega_m+\Omega_\Lambda +\Omega_k  =1+\xi-\frac{\omega}{6}\xi^2
 \end{equation}
\begin{equation}\label{eq14}
\Omega_{\Lambda}= \frac{\omega+3}{2\omega}\Omega_m -\frac{2\omega+3}{2\omega}q
+ \frac{2\omega+3}{6}\xi^2 - \frac {2\omega+3}{2\omega}\xi
\end{equation}
\begin{equation}\label{eq15}
\Omega_{k}=1 - \frac{3(\omega+1)}{2\omega}\Omega_m +\frac{2\omega+3}{2\omega}q
+ \frac{4\omega+3}{2\omega}\xi - \frac {\omega+1}{2}\xi^2
\end{equation}
\begin{equation}\label{eq16}
\Omega_{m}= q-\frac{\omega}{3}q_{\phi}+(\omega+1)\xi-\frac{\omega}{3}\xi^2
\end{equation}
where,
\begin{equation}\label{eq17}
\xi= \frac{\dot{\phi}}{\phi H}\; \&\; q_{\phi}=-\frac{\ddot{\phi}}{\phi   H^2}
\end{equation}
The Eq.(\ref{eq13}) is the Brans-Dicke analogue of density parameter relation ship of the CDM relativistic model.
\subsection{Spatially flat dust model }
We consider spatially flat space  ($k=0,\Omega_k=0 $).  The Eqs.(\ref{eq15}) and (\ref{eq16}) give
rise to the following equation.
\begin{equation}\label{eq18}
q-(\omega+1)q_{\phi}+(3\omega+2)\xi=2
\end{equation}
 The Eq.(\ref{eq18}) has first integral
\begin{equation}\label{eq19}
(\omega+1)\frac{\dot{\phi}}{\phi}-\frac{\dot{a}}{a}=\frac{L}{\phi a^3}
\end{equation}
Where $L$ is constant of integration.
\section{Gravitational constant versus redshift relation }
The solution Eq.(\ref{eq19}) has a singularity at $a=0\ and \ \phi=0$, so we take  constant $L=0$. This gives the  following power law relation between scalar field $\phi$
and scale factor$\;a$.
\begin{equation}\label{eq20}
\xi= \frac{1}{\omega+1},~~\phi=\phi_0\left(\frac{a}{a_0}\right)^{\frac{1}{\omega+1}}
\end{equation}
Where $\phi_0$ and $a_0$ are values of scalar field $\phi$ and scale factors $a$ at present.
As gravitational constant $G$ is reciprocal of $\phi$ i.e.   $$G=\frac{1}{\phi}\;,$$
and $$\frac{a_0}{a}=(1+z),$$ Where $z$ is the red shift.\\
So, $$\frac{{G} }{G_0}= \left(1+z\right)^{\frac{1}{\omega+1}}$$
This relation ship shows that  $G$ grows toward the past and in fact it diverges at cosmological singularity. Radar observations, Lunar mean motion and the  Viking landers on Mars (Narlikar~ \cite{ref40}) suggest that rate of variation of gravitational constant must be very much slow. The recent experimental evidence shows that $\omega>40000$ (Bertotti, B., et al.~\cite{ref41}; Felice, A.D. et al.~\cite{ref42}). Accordingly, we consider large  Coupling constant
\begin{equation}\label{eq21}
  \omega=40000
\end{equation}
 From (\ref{eq19}), the present rate of  gravitational constant is given by
\begin{equation}\label{22}
\left(\frac{\dot{G}}{G}\right)_0  = -\frac{1}{\omega+1} H_0=2.5\times 10^{-15}
\end{equation}
 where we have taken $ H_0 \simeq  10^{-10} year^{-1}$.\;
 Fig1 exhibits the fact that how \;$G/G_0$ varies over $\omega$.\; For higher values of $\omega$, \; $G/G_0$  grows very slow over redshift, where as  for lower values of $\omega$ it grows fast.\\
  \begin{figure}
   \centering
   \includegraphics[width=0.75\textwidth, angle=0]{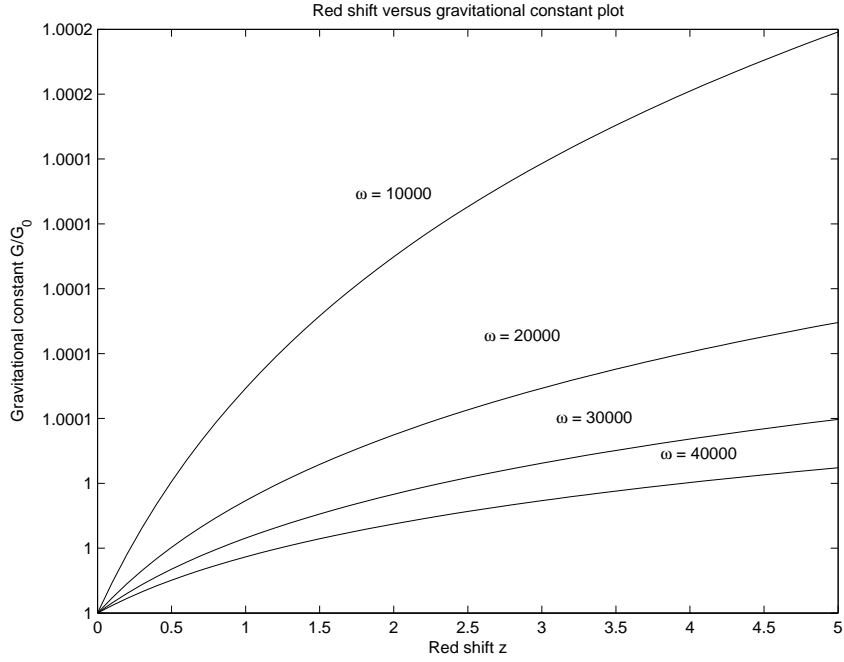}
   \caption {Variation of Gravitatation constant over red shift for different $\omega$'s   }
   \label{Fig1}
   \end{figure}
\subsection{Density parameters},
Eqs. (\ref{eq13}) and (\ref{eq20}) give
\begin{equation}\label{eq23}
\Omega_m+\Omega_\Lambda=1+\frac{5\omega+6}{6(\omega+1)^2}
\end{equation}
For $\omega=40000$,~  Eq (\ref{eq23}) becomes
 \begin{equation}\label{eq24}
 \Omega_m+\Omega_\Lambda=1.0000208
 \end{equation}
 \subsection{Expression for Hubble's constant}
 Integration of energy conservation  Eq.(\ref{eq10}) gives
 \begin{equation}\label{eq25}
 \rho=(\rho)_0\left(\dfrac{a_0}{a}\right)^3
 \end{equation}
 Where, we have taken $\gamma=1$ for dust matter.\\
 Eqs (\ref{eq12}),(\ref{eq20}), (\ref{eq23}) and (\ref{eq25}) give expression for Hubble's constant as
 \begin{equation}\label{eq26}
 H=\frac{H_0}{\sqrt{1+\frac{5\omega+6}{6(\omega+1)^2}}}
 \sqrt{(\Omega_m)_0 \left(\frac{a_0}{a}\right)^{\frac{3\omega+4}{\omega+1}}+(\Omega_{\Lambda})_0}
 \end{equation}
 As $$\frac{a_0}{a}=(1+z),$$  Hubble's constant in term of red shift is given by
 \begin{equation}\label{eq27}
 H=\frac{H_0}{\sqrt{1+\frac{5\omega+6}{6(\omega+1)^2}}}
 \sqrt{(\Omega_m)_0 (1+z)^{\frac{3\omega+4}{\omega+1}}+(\Omega_{\Lambda})_0}
 \end{equation}
 \section{Expression for Luminosity Distance }
  The luminosity distance in metric (\ref{eq4}) is as follows
  $$D_{L}=a_{0}r(1+z)$$
  To get expression for luminosity distance we consider light
  travailing along radial direction $r$. It satisfies null geodesic
  given by
  $$ds^{2}=c^{2}dt^{2}-a(t)^{2}dr^{2}=0$$
  From this, we obtain
  $$r=\int_{0}^{r_{s}}dr=\int_{t_{1}}^{t_{0}}\frac{dt}{a(t)}=\int_{z}^{0}\frac{dz}{a\dot{z}} =\frac{1}{a_{0}}\int_{0}^{z}\frac{dz}{H(z)}$$
  Where,
  $$\dot{z}=\dot{\left(\frac{a_{0}}{a}\right)}=-H\left(\frac{a_0}{a}\right)$$
  So the  luminosity distance  $D_L$ is given by
  \begin{equation}\label{eq28}
  D_{L}=\frac{c(1+z)\sqrt{1+\frac{5\omega+6}{6(\omega+1)^2}}}{H_{0}}\int_{0}^{z}\frac{dz}{\sqrt{[(\Omega_{m})_{0}(1+z)^{\frac{3\omega+4}{\omega+1}}+(\Omega_{\Lambda})_{0}]}}\,
  \end{equation}
  \subsection{Expression for Apparent Magnitude}
  The apparent magnitude of a source of light is related to the  luminosity distance via following  expression
  \begin{equation}\label{eq29}
  m = 16.08 + 5 log_{10}\frac{H_{0}D_{L}}{.026c Mpc}
  \end{equation}
  \begin{equation}\label{eq30}
  log_{10}(\frac{H_{0}D_{L}}{c})= (m - 16.08)/5 + log_{10}(.026 Mpc)
  \end{equation}
  Using (\ref{eq28}),  we get following expression for apparent magnitude
  \begin{equation}\label{eq31}
  m=16.08+5log_{10}\left(\frac{(1+z)\sqrt{1+\frac{5\omega+6}{6(\omega+1)^2}}}{.026}\int_{0}^{z}\frac{dz}{\sqrt{[(\Omega_{m})_{0}(1+z)^{\frac{3\omega+4}{\omega+1}}+(\Omega_{\Lambda})_{0}]}}\right)
  \end{equation}
  \section{ Estimation of present values of energy parameters}
  We consider $287$ high red shift ($ 0.3 \leq z \leq 1.4$ ) SN Ia supernova data set of observed apparent
  magnitudes along with their possible error from union $2.1$ compilation (Suzuki,N. et al.~ \cite{ref43}).  We also obtain a large number of
  theoretical data set corresponding to $(\Omega_m)_0$ in the range
  ($ 0 \leq (\Omega_m)_0 \leq 1$ ) from equations (\ref{eq21}),~(\ref{eq24})~and~(\ref{eq31}).\\

  In order to get the best fit theoretical data set of apparent magnitudes, we calculate
  $\chi^{2}$  by using following statistical formula (Yadav et al.~\cite{ref44}).
  \begin{equation}\label{eq32}
  \chi_{SN}^{2}= \frac{A-\frac{B^{2}}{C}+log_{10}\left(\frac{C}{2\pi}\right)}{287},
  \end{equation}
  where,
  \begin{equation}\label{eq33}
  A=\overset{287}{\underset{i=1}{\sum}}\frac{\left[\left(m\right)_{ob}-\left(m\right)_{th}\right]^{2}}
  {\sigma_{i}^{2}},
  \end{equation}
  \begin{equation}\label{eq34}
  B=\overset{287}{\underset{i=1}{\sum}}\frac{\left[\left(m\right)_{ob}-\left(m\right)_{th}\right]}
  {\sigma_{i}^{2}},
  \end{equation}
  and
  \begin{equation}\label{eq35}
  C=\overset{287}{\underset{i=1}{\sum}}\frac{1}{\sigma_{i}^{2}}.
  \end{equation}
  Here the sums are taken over data sets of observed and theoretical values of apparent magnitudes of $287$ supernovae. \\
  Using Eqs.  (\ref{eq32})-(\ref{eq35}),  we find that for minimum value of $\chi^2=16.6910$,  the best fit present values of $\Omega_{m}$ and
  $\Omega_{\Lambda}$ are given by $ (\Omega_{m})_0 = 0.296 $ and $ (\Omega_{\Lambda})_0 = 0.712 $.\\
  Now we repeat the above process with luminosity distance.
  The observed data set of luminosity distances  are obtained from those of  apparent
  magnitude data set given in the union $2.1$ compilation by using equation (\ref{eq28}). We get a large number of
  data sets of theoretical values of luminosity distances corresponding to $(\Omega_m)_0$ in the
  range ($ 0 \leq (\Omega_m)_0 \leq 1$ ) from equation (\ref{eq21}),~(\ref{eq24})~and~ (\ref{eq28}). We find that for minimum value of $\chi^2=0.6545$,  the best fit present values of $\Omega_{m}$ and $\Omega_{\Lambda}$ is  presented  again as $ (\Omega_{m})_0 = 0.296 $ and $ (\Omega_{\Lambda})_0 = 0.704 $.
  The Figures $2$ and $3$ also indicates how the observed values of apparent magnitudes and luminosity distances
  reach close to the  theoretical graphs for $(\Omega_\Lambda)_0$ = 0.704 and $(\Omega_m)_0 = 0.296$.
  \newpage
  \begin{figure}
  	\begin{minipage}{0.50\textwidth}
  		\centering
  		\includegraphics[width=.9\textwidth]{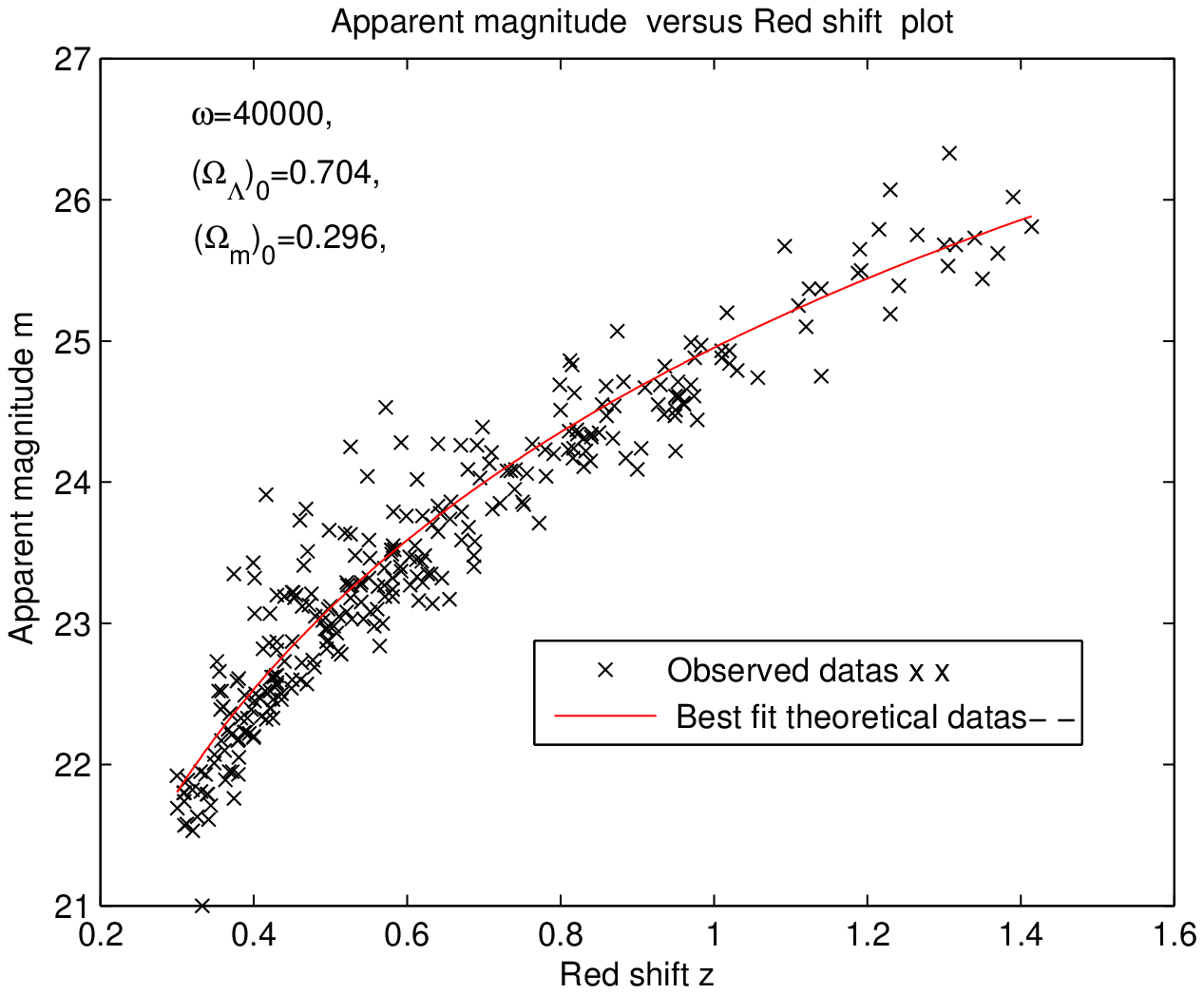}
  		\caption{Apprent magnitude versus red shift best fit curve}
  	\end{minipage}
  	\hfill
  	\begin{minipage}{0.50\textwidth}
  		\centering
  		\includegraphics[width=.9\textwidth]{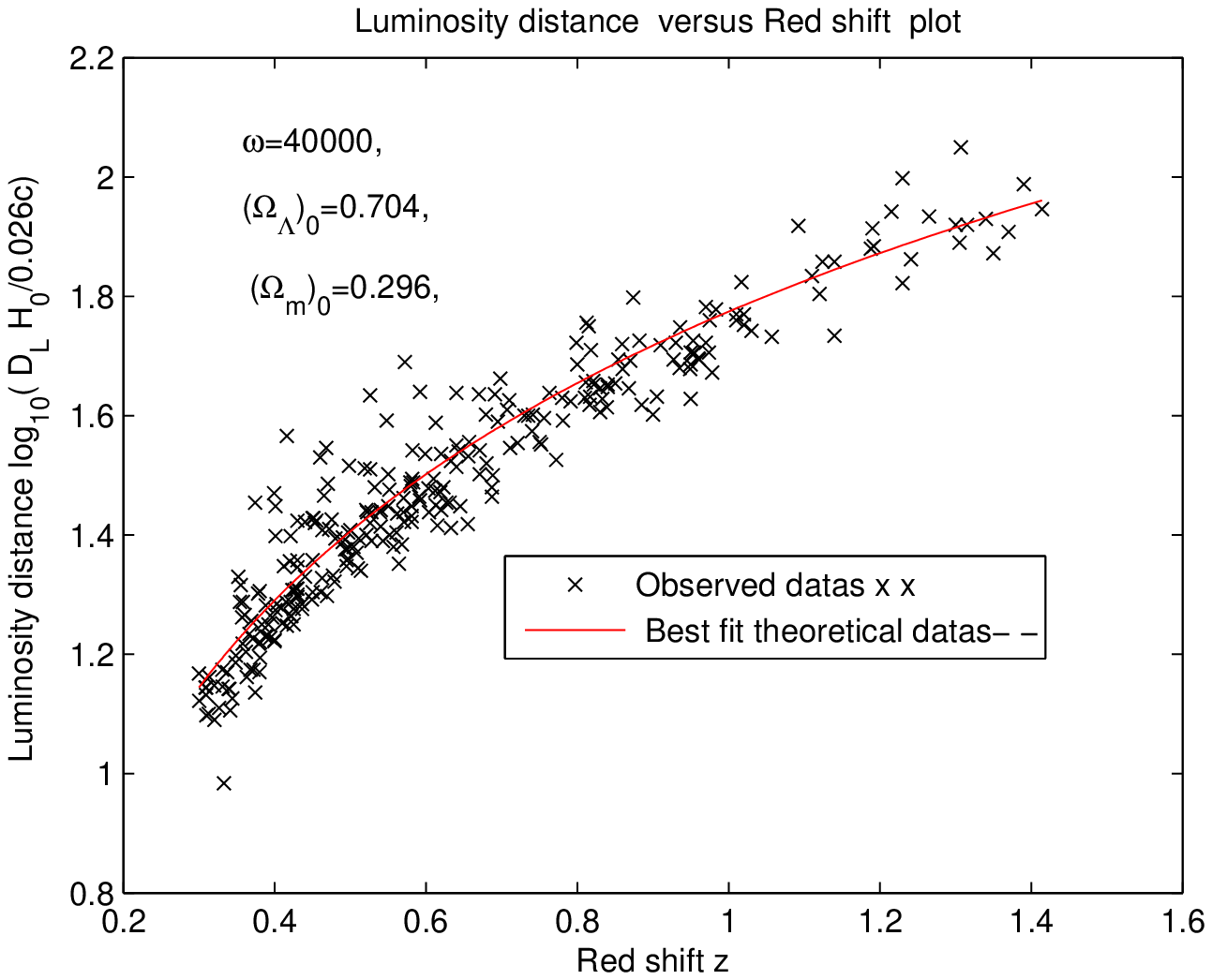}
  		\caption{ Luminosity distance versus red shift best fit curve }
  	\end{minipage}
  \end{figure}
  \section{Estimation of present values of Hubble's constant $H_0$}
  We present a data set of the observed values of the Hubble parameters H(z) versus the red shift z with possible
  error in the form of  Table-1. These data points were obtained by various researchers from time to time,
  by using differential age approach.\\
\begin{table}
  \begin{center}
\caption[]{ Hublle's constant Table }
  	\begin{tabular}{|c|c|c|c|c|}
  		\hline $z$ & $H(z)$ &
  		$\sigma_{H}$ & Reference & Method \tabularnewline \hline \hline
  		0.07	&	69	&	19.6	&	 Moresco M. et al., 2012	&	DA	\tabularnewline
  		\hline									
  		0.1	&	69	&	12	&	 Zhang C. at el., 2014	&	DA	\tabularnewline
  		\hline									
  		0.12	&	68.6	&	26.2	&	 Moresco M. et al., 2012	&	DA	\tabularnewline
  		\hline									
  		0.17	&	83	&	8	&	 Zhang C. at el., 2014	&	DA	\tabularnewline
  		\hline
  		0.28	&	88.8	&	36.6	&	 Moresco M. et al., 2012	&	DA	\tabularnewline
  		\hline									
  		0.4	&	95	&	17	&	 Zhang C. at el., 2014	&	DA	\tabularnewline
  		\hline									
  		0.48	&	97	&	62	&	 Zhang C. at el., 2014	&	DA	\tabularnewline
  		\hline									
  		0.593	&	104	&	13	&	 Moresco M., 2015	&	DA	\tabularnewline
  		\hline									
  		0.781	&	105	&	12	&	 Moresco M., 2015	&	DA	\tabularnewline
  		\hline									
  		0.875	&	125	&	17	&	 Moresco M., 2015	&	DA	\tabularnewline
  		\hline									
  		0.88	&	90	&	40	&	 Zhang C. at el., 2014	&	DA	\tabularnewline
  		\hline									
  		0.9	&	117	&	23	&	 Zhang C. at el., 2014	&	DA	\tabularnewline
  		\hline									
  		1.037	&	154	&	20	&	 Moresco M., 2015	&	DA	\tabularnewline
  		\hline									
  		1.3	&	168	&	17	&	 Zhang C. at el., 2014	&	DA	\tabularnewline
  		\hline									
  		1.363	&	160	&	33.6	&	 Moresco M., 2015	&	DA	\tabularnewline
  		\hline									
  		1.43	&	177	&	18	&	 Zhang C. at el., 2014	&	DA	\tabularnewline
  		\hline									
  		1.53	&	140	&	14	&	 Zhang C. at el., 2014	&	DA	\tabularnewline
  		\hline									
  		1.75	&	202	&	40	&	 Zhang C. at el., 2014	&	DA	\tabularnewline
  		\hline									
  		1.965	&	186.5	&	50.4	&	  Stern D at el.,2010	&	DA	\tabularnewline
  		\hline									
  	\end{tabular}
  	\end{center}
    \end{table}

  As per our model,  Hubble's constant H(z) versus red shift 'z' relation
  is given by Eq. (\ref{eq27} ). Hubble Space Telescope (HST) observations of
   Cepheid variables \cite{ref49} provides present value of Hubble's constant in the range $H_0 = 73.8 \pm 2.4 km/s/Mpc$ .
  Taking $\omega=40000, (\Omega_{m})_{0}=0.296, (\Omega_{\Lambda})_{0}=0.704$ and using equation (\ref{eq27}), a large number of  data sets of theoretical values of Hubble's constant H(z) for red shifts as per table-1 and $H_0$
  in the range ($ 69 \leq H_0 \leq 74$ )  are obtained. It should be noted that each
  data set will consist of $19$ data points and data sets differ due to changing values of $H_0$ \\

  In order to get the best fit theoretical data set of Hubble's constant $H(z)$ versus $z$, we calculate
  $\chi^{2}$  by using following statistical formula.
  \begin{equation}\label{eq36}
  \chi_{SN}^{2}= \overset{19}{\underset{i=1}{\sum}}\frac{\left[\left(H\right)_{ob}-\left(H\right)_{th}\right]^{2}}
  {\sigma_{i}^{2}},
  \end{equation}
  Using Eq. (\ref{eq36}),  we find that for minimum value of $\chi^2 =10.2558$, the best fit present value of Hubble's constant
  $H_0$ is $72.30$ /s/Mpc. From Figures $4$ and $5$ we also observe  the dependence of Hubble's constant with
  red shift and scale factors. In the figure $4$, Hubble's observed data points are closed to the graph corresponding
  to $(\Omega_\Lambda)_0$ = $0.704$, $(\Omega_m)_0 = 0.296$ and $H_0 = 72.30$ /s/Mpc . This validates the proximity of observed and theoretical
  values.
  \begin{figure}
   	\begin{minipage}{.45\textwidth}
   		\centering
   		\includegraphics[width=.9\textwidth]{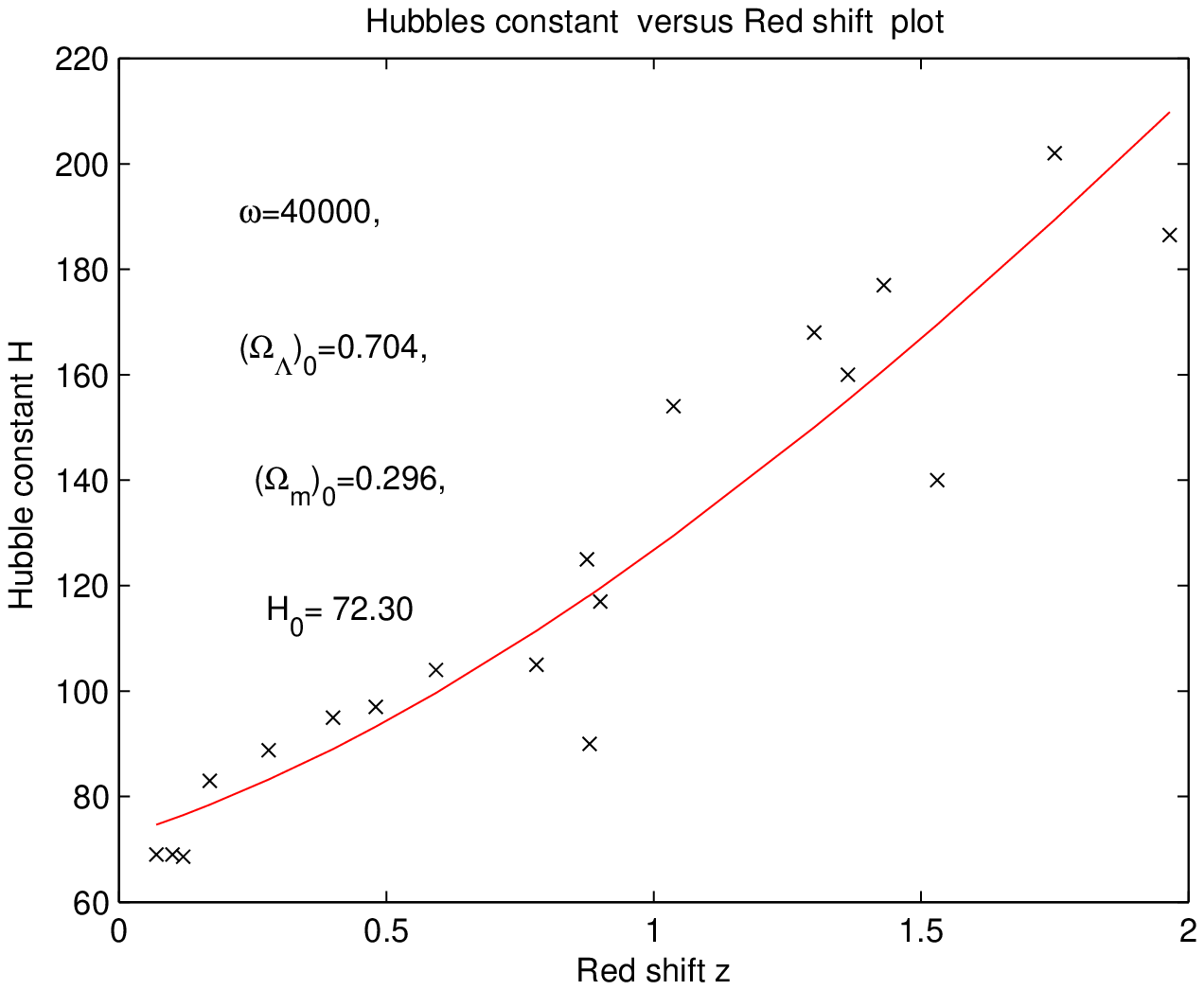}
   		\caption{Variation of Hubble's constant over red shift;  Best fit curve}
   	\end{minipage}
   	\hfill
   	\begin{minipage}{.45\textwidth}
   		\centering
   		\includegraphics[width=.9\textwidth]{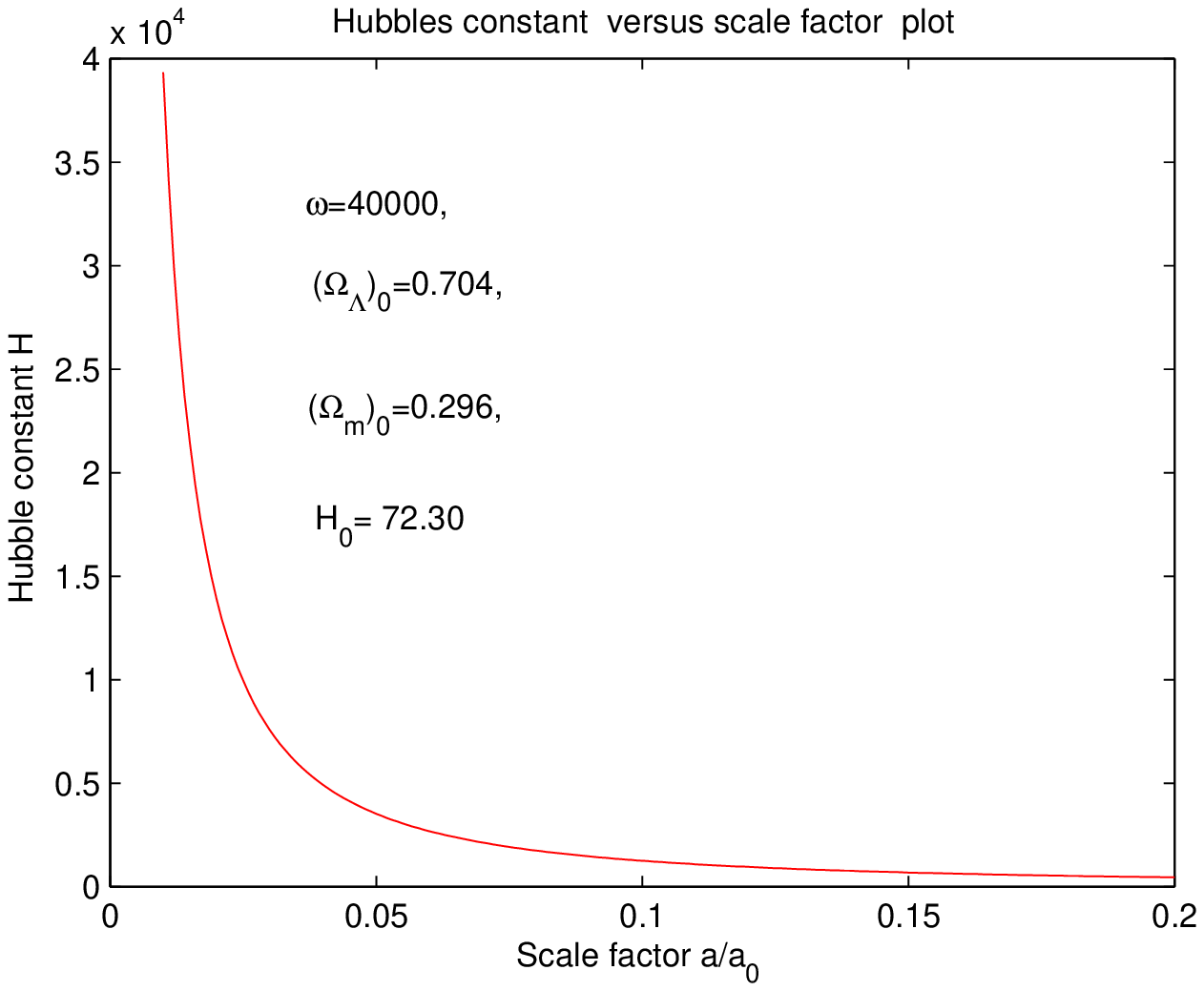}
   		\caption{Variation of Hubble's constant over scale factor}
   	\end{minipage}
   \end{figure}
  \section{Certain physical properties of the universe}
  \subsection{Matter and dark energy densities  }
  The matter and  dark energy densities of the universe are related to the energy parameters through following equation
  \begin{equation}\label{eq37}
  \Omega_{m}=\frac{(\rho)_m}{\rho_{c}},\Omega_{\Lambda}=\frac{\rho_{\Lambda}}{\rho_{c}}
  \end{equation}
  Where
  \begin{equation}\label{eq38}
  \rho_{c}=\frac{3c^{2}H^{2}}{8\pi G}=\frac{3c^{2}\phi H^{2}}{8\pi }
  \end{equation}
 So,
  \begin{equation}\label{eq39}
   (\rho_m)_0=(\rho_{c})_0 (\Omega_{m})_0= ,\; (\rho_{\Lambda})_0=(\rho_{c})_0(\Omega_{\Lambda})_0
  \end{equation}
  Now $$  (\rho_{c})_0=\frac{3c^{2}H_{0}^{2}}{8\pi G} = 1.88\; h^2_0 \times 10^{-29}\; gm/cm^3$$
  Therefore, the present value of matter  and dark energy densities are given by
  \begin{equation}\label{eq40}
  (\rho_{m})_{0}= 0.5565 h^2_0\times10^{-29}gm/cm^{3}.
  \end{equation}
  And
   \begin{equation}\label{eq41 }
   (\rho_{\Lambda})_{0}=\rho_c(\Omega_{\Lambda})_{0} = 1.3235 h^2_0\times10^{-29}gm/cm^{3}.
   \end{equation}
   Where we have taken $$ (\Omega_{m})_0= 0.296\; \& \;  (\Omega_{m})_0= 0.704  $$
   General expressions for matter and dark energies are given by
   \begin{equation}\label{eq42}
   \rho=(\rho)_0\left(\dfrac{a_0}{a}\right)^3=(\rho)_0\left(1+z\right)^3
   \end{equation}
 And
  \begin{equation}\label{eq43}
  (\rho_{\Lambda})=\rho_c\;\Omega_{\Lambda}.
  \end{equation}

  We see that the current matter and dark energy densities are very close to the values predicted by
  the various surveys described in the introduction.

    \subsection{Age of the universe}
    We begin with the integral
    \begin{equation*}
    t = \intop_{0}^{t}dt=\intop_{0}^{a}\frac{da}{aH}
    \end{equation*}
    \begin{equation}\label{eq44}
    t  =  \intop_{0}^{a}\frac{\sqrt{1+\frac{5\omega+6}{6(\omega+1)^2}}da}{aH_0
    	\sqrt{(\Omega_m)_0 (\frac{a_0}{a})^{\frac{3\omega+4}{\omega+1}}+(\Omega_{\Lambda})_0}}
    \end{equation}
    Integrating Eq.(\ref{eq44}), we get the following expression for age of the universe.
    \begin{equation}\label{eq45}
    H_{0}t=\frac{2\sqrt{1+\frac{5\omega+6}{6(\omega+1)^2}}}{\frac{3\omega+4}{\omega+1}\sqrt{(\Omega_{\Lambda})_{0}}}
    log\left(\sqrt{\frac{(\Omega_{\Lambda})_{0}}{(\Omega_{m})_{0}} \left(\frac{a}{a_0}\right)^{\frac{3\omega+4}{\omega+1}}}
    +\sqrt{1+\frac{(\Omega_{\Lambda})_{0}}{(\Omega_{m})_{0}} \left(\frac{a}{a_0}\right)^{\frac{3\omega+4}{\omega+1}}}\right)
    \end{equation}
    In term of red shift, the  age is given as,
    \begin{equation}\label{eq46}
    H_{0}t=\frac{2\sqrt{1+\frac{5\omega+6}{6(\omega+1)^2}}}{\frac{3\omega+4}{\omega+1}\sqrt{(\Omega_{\Lambda})_{0}}}
    log \left (\sqrt{\frac{(\Omega_{\Lambda})_{0}}{(\Omega_{m})_{0}} \left(\frac{1}{1+z}\right)^{\frac{3\omega+4}{\omega+1}}}
    +\sqrt{1+\frac{(\Omega_{\Lambda})_{0}}{(\Omega_{m})_{0}} \left(\frac{1}{1+z}\right)^{\frac{3\omega+4}{\omega+1}}}  \right)
    \end{equation}
    The present age of the universe is given by
    \begin{equation}\label{eq47}
    H_{0}t_0=\frac{2\sqrt{1+\frac{5\omega+6}{6(\omega+1)^2}}}{\frac{3\omega+4}{\omega+1}\sqrt{(\Omega_{\Lambda})_{0}}}
    log\left( \sqrt{\frac{(\Omega_{\Lambda})_{0}}{(\Omega_{m})_{0}}} +\sqrt{1+\frac{(\Omega_{\Lambda})_{0}}{(\Omega_{m})_{0}}}\right)
    \end{equation}
    For $\omega=40000$, $(\Omega_{\Lambda})_{0}$=$0.704$ and $(\Omega_{m})_{0}= 0.296$, $H_{0}t_{0}=0.9677$ .
    Since $ H_{0}^{-1}=9.776h^-1$ Gyr = 13.5778 Gyr when $h=0.723 $. Therefore $ t_{0}$ = $13.0847$ Gyr.
    This is consistence with most recent WMAP data $t_{0}=13.73_{-0.17}^{+0.13}$ . \\
    In Figures $6$ and $7$, we have shown the variation of time over  scale factor and red shift. This also indicated the consistency with recent observations.
    \begin{figure}
    	\begin{minipage}{.50\textwidth}
    		\centering
    		\includegraphics[width=.9\textwidth]{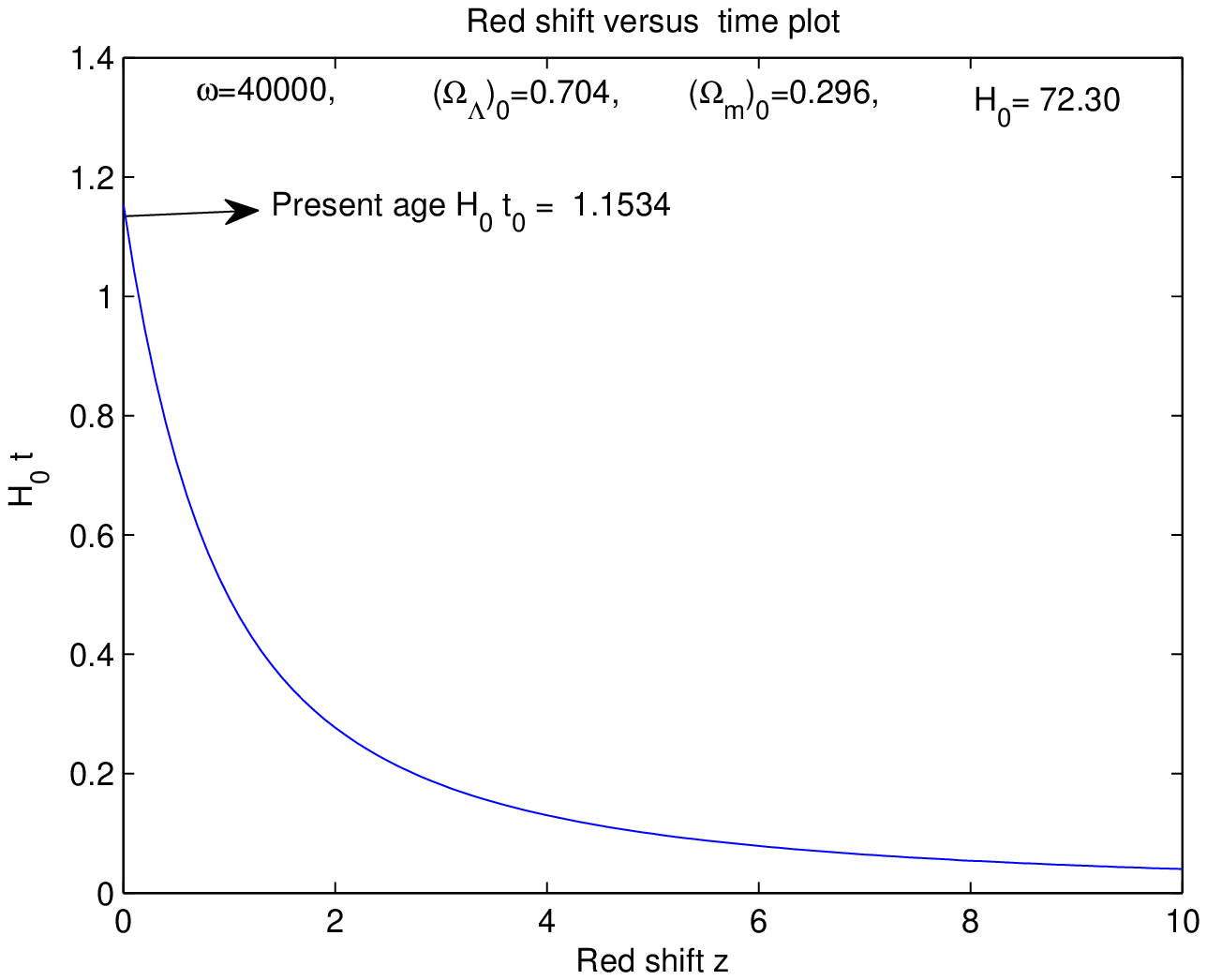}
    		\caption{Variation of red shift over time}
    	\end{minipage}
    	\hfill
    	\begin{minipage}{.50\textwidth}
    		\centering
    		\includegraphics[width=.9\textwidth]{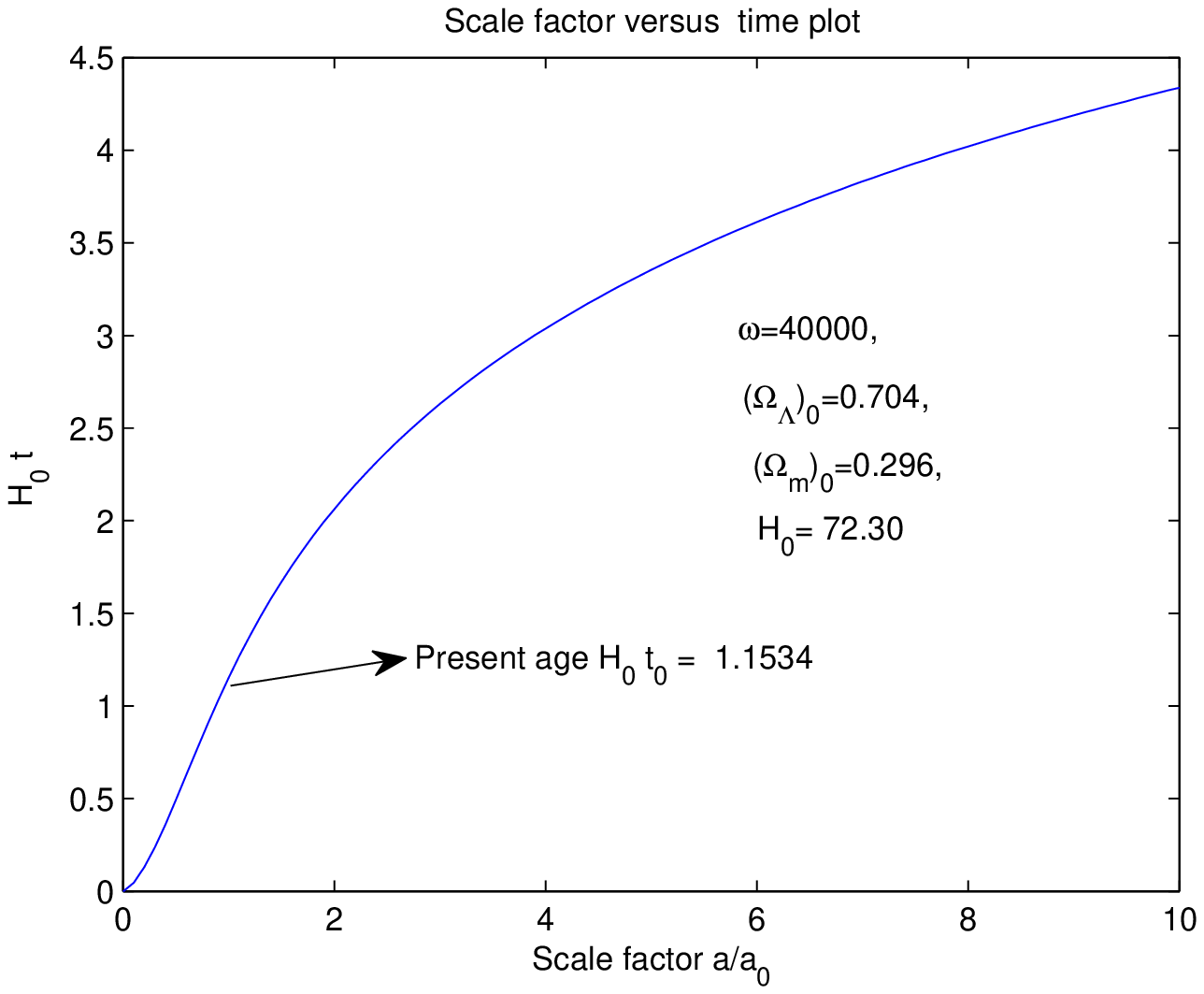}
    		\caption{Variation of scale factor over time }
    	\end{minipage}
    \end{figure}

\subsection{  Deceleration parameter }
  Eqs.(\ref{eq14}), (\ref{eq20}) and (\ref{eq23}) give
\begin{equation}\label{eq48}
q =\frac{\omega+2}{2(\omega+1)}- \frac{3(\omega+1)}{2\omega+3}\;\Omega_\Lambda
\end{equation}
Using Eqs.(\ref{eq26}) and (\ref{eq27}), we get following expression for
deceleration parameter
\begin{equation}\label{eq49}
q =\frac{\omega+2}{2(\omega+1)} - \frac{3(\omega+1)}{2\omega+3}\;\left(\Omega_\Lambda \right)_0\frac{\left(1+\frac{5\omega+6}{6(\omega+1)^2}\right)}
{\left((\Omega_m)_0\left(\frac{a_0}{a}\right)^{\frac{3\omega+4}{\omega+1}}+(\Omega_{\Lambda})_0\right)}
\end{equation}
In term of redshift it is given by
\begin{equation}\label{eq50}
q =\frac{\omega+2}{2(\omega+1)} - \frac{3(\omega+1)}{2\omega+3}\;\left(\Omega_\Lambda \right)_0\frac{\left(1+\frac{5\omega+6}{6(\omega+1)^2}\right)}
{\left((\Omega_m)_0\left(1+z\right)^{\frac{3\omega+4}{\omega+1}}+(\Omega_{\Lambda})_0\right)}
\end{equation}

As the present phase ($z=0$) of the universe is accelerating $q \leq 0\; ie\; \frac{\ddot{a}}{a}\geq 0 $ , so
we must have
\begin{equation}\label{eq51}
(\Omega_\Lambda)_0 \geq\dfrac{(2\omega+3)(\omega+2)}{6(\omega+1)^2}
\end{equation}

For $\omega=40000$, the limit is as follows
$$(\Omega_\Lambda)_0 \geq 0.3333$$
which is consistent with the present observed value of $(\Omega_{\Lambda})_{0}=0.704$.
Putting $\omega=40000$, $(\Omega_{\Lambda})_{0}$=$0.704$, $(\Omega_{m})_{0}= 0.296$ and $z=0$ in Eq.(\ref{eq50}), we get the present value of deceleration parameter as
\begin{equation}\label{52}
q_0 = -0.5560
\end{equation}
The universe attains to the accelerating phase when $z<z_c$ where $z=z_c\; at\; q=0$ .  The Eq.(\ref{eq50}) provides
\begin{equation}\label{53}
z_{c} \cong 0.6818
\end{equation}
 Thus, the acceleration must have begun in the past at
 $z_{c}=0.6818 \thicksim .5180 H^{-1}_0 yrs\thicksim 7.2371\times10^9 yrs$ before from present. We have converted red shift by time from Eq.(\ref{eq46}). The figure (8) Shows how deceleration parameter increases from negative to positive over redw shift which means that in the past universe was decelerating, then at a instant $z_{c} \cong 0.6797$ it became stationary there after it starts accelerating.
\begin{figure}
	\centering
	\includegraphics[width=.75\textwidth]{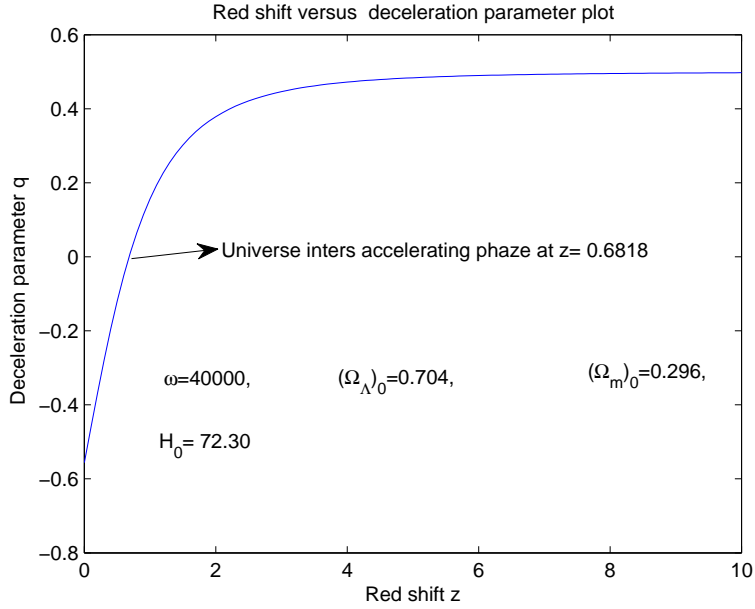}	
	\caption{Variation of decelerating parameter over red shift. It represents accelerating universe at present.  }	
\end{figure}
 \section{Conclusion}
 We summarize our results by presenting  Table-2  which displays the values of cosmological parameters at
 present obtained by us.
\begin{table}
 \begin{center}
\caption[]{Cosmological  parameters at present}
 	\begin{tabular}{|c|c|}
 		\hline
 		Cosmological Parameters & Values at Present \\
 		\hline
 		BD coupling constant $\omega$ & 40000\\
        $\left(\frac{\dot{G}}{G}\right)_0$ & 2.5$\times 10^{-15}$\\
		 Dark energy parameter $(\Omega_\Lambda)_0$ & 0.704 \\
 		 Dust energy parameter $(\Omega_m)_0$ & 0.296\\
 		 Hubble's constant $H_0$ & 72.30\\
 		Deceleration parameter $(q)_0$& -0.5560\\
 		Dust energy density $(\rho_{m})_{0}$ &$ 0.5565h^2_0\times10^{-29}gm/cm^{3}$  \\
 		Dark energy density $(\rho_{\Lambda})_{0}$& $ 1.3235h^2_0\times10^{-29}gm/cm^{3}$\\
 		 Age of the universe $t_{0}$&$13.0847 Gyr$\\
 		\hline
 	\end{tabular}
 \end{center}
\end{table}

 We have found  that the acceleration would have begun in the past at
 $z_{c}=0.6818 \thicksim 7.2371\times10^9 yrs$ before from present.
 These results are in good agreements with the various surveys described in the introduction.
 \section*{Acknowledgments}
This work is  supported by the CGCOST  Research Project
789/CGCOST/MRP/14. The author is  thankful to IUCAA, Pune, India for providing facility
and support where part of this work was carried out during a visit. Author is also thankful to Prof J V Narlikar, IUCAA , Prof DRK Reddy, Andhra University and Prof Anirudh Pradhan, G. L. A. University for seeing the paper and making useful comment.

\label{lastpage}
\end{document}